# *Retrieval of depth profile of nano scale thin films by one directional polarization analysis in neutron specular reflectometry*

S. Farhad Masoudi[*], Saeed S. Jahromi

Department of Physics, K.N. Toosi University of Technology, P.O. Box 15875-4416, Tehran, Iran

**Abstract**

Recently it has been shown that the modules and phase of complex reflection coefficient can be determined by using a magnetic substrate and polarized neutrons. Several other methods have also been worked out based on measurement of polarizations of reflected neutrons from magnetic reference layers and magnetic substrate. However, due to the fact that available reflectometers are limited in the choice of polarization of reflected beam in the same direction as the polarization of the incident beam, neither of the methods which are based on polarization analysis, has been proven to be experimentally practical. In this paper, we have proposed a new method for determining the phase of reflection coefficient which is based on two measurements of polarization which correspond to two magnetic fields with the same magnitudes and different orientations. The polarization analysis is performed in the same direction as the polarization of the incident beam and is well suited for available reflectometers. The problems envisaged in implementation of the method are also discussed.

*Corresponding author.
E-mail: masoudi@kntu.ac.ir

## 1. Introduction

Measurement of intensity of neutrons reflected elastically and specularly from thin films, as function of the glancing angle of incident, provides us with useful information about the depth profile of the sample [1]. However, extracting the profile from the measured reflectivity, $|r(q)|^2$, as a function of wave vector $q$ is difficult and no direct inversion scheme is possible in the absence of the full knowledge of reflection coefficient, r(q) (modules and phase) [1,2]. Due to the importance of phase information to the study of thin films structure, finding a practical solution to this so-called phase problem is of great interest. Recently, several experimental and theoretical efforts have been devoted to finding the knowledge of phase [2-14] among which, the reference method and variation of surroundings are of highest applicability. The reference layer method is based on the interference between the reflections of a known reference layer and the unknown surface profile [3,4] and the method of variation of surroundings medium, make use of the controlled variations of scattering length density of the incident and/or substrate medium instead of reference layers of finite thickness [5,6]. Although these methods are highly promising, only the method of variation of substrates has been experimentally proven to be practical [5]. Obviously, changing the surrounding medium is the most challenging part of the experiment. Majkrzak et al. proposed using a gas-liquid or solid-liquid interface and extracted the phase information by using $H_2O$ and $D_2O$ as substrate [5,6].

At present, several other promising methods have also been worked out which make use of the spin-dependent interaction of a neutron with a magnetic reference layer [7–10] or magnetic substrate [11,12]. In a more recent research, Leeb et. al, have developed the reference method by using polarization of the reflected neutron and left and right reflection coefficients of the reference layer as known parameters [8]. Correspondingly, we have developed Leeb's method of magnetic reference layer with polarization analysis in a straightforward manner by proposing a change into the placement of the known and unknown layers [9,10]. We have also developed the method of variation of surroundings, by using a magnetic substrate and polarization analysis [11,12].

In those methods which are based on polarization analysis, generally we have to measure the polarization of the reflected neutrons in different directions as the polarization direction of the incident neutron; beside this fact that more than one measurement of polarization is required. However, available reflectometers are limited with measurement of polarization of reflected neutrons in the same direction as polarization direction of the incident neutrons.

Recently, Leeb et. al have worked out a method in which they considered this limitation by polarization analysis in two different directions [13]. The problem is solved by changing the coordinate's direction by rotating the sample and measuring the polarization for each case. In this paper, we have solved this problem by using a magnetic substrate as reference layer that make the method more simple. The method is based on polarization analysis of reflected beam in the same direction as the incident beam, without any need to reflectivity data.

## 2. Method

Consider an unknown layer which is mounted on top of a magnetic substrate, oriented in the following right-handed coordinate system: $x$, is the depth direction, $z$ is along the direction of the magnetization in the magnetic substrate and $y$ is along the normal vector to the plane of reflection (Figure 1). The scattering length density of the magnetic substrate is proportional to $(\rho(x) \pm \rho_m)$ where, $\rho_m = \frac{m}{2\pi\hbar^2}\mu B$ is the magnetic SLD of the sample, $\rho$ is the SLD at depth x from the surface, $B$ is the magnetic field, and $\mu$ is magnetic moment of neutrons. The two signs for the second term refer to neutrons polarized parallel and anti parallel to the local magnetization, respectively [1,2].

Here we use the transfer matrix method that makes the method simpler than that of Ref [13]. For a sample with thickness L mounted on top of a magnetic substrate, the transmission and reflection coefficients, $t_\pm$ and $r_\pm$ respectively, can be calculated exactly from the 2×2 transfer matrix which carries the exact wave function and its first derivative across the film, from front edge to back:

$$\begin{pmatrix} 1 \\ ih_\pm \end{pmatrix} t_\pm e^{iqL} = \begin{pmatrix} A(q) & B(q) \\ C(q) & D(q) \end{pmatrix} \begin{pmatrix} 1+r_\pm \\ if(1-r_\pm) \end{pmatrix},$$

(1)

where the elements of transfer matrix, $A$, $B$, $C$ and $D$, are uniquely determined by the SLD of the sample and are real when there is no effective absorption. $h_\pm$, is the refractive index of the magnetic substrate which corresponds with plus and minus magnetization of the substrate as follows:

$$h_\pm = \left(1 - \frac{4\pi(\rho \pm \rho_m)}{q^2}\right)^{1/2}$$

(2)

Using equation (1) we have [2]:

$$r_\pm(q) = \frac{\beta_\pm^{fh} - \alpha_\pm^{fh} - 2i\gamma_\pm^{fh}}{\beta_\pm^{fh} + \alpha_\pm^{fh} + 2}$$

(3)

where

$$\alpha_\pm^{fh} = h_\pm f^{-1} A^2 + (fh_\pm)^{-1} C^2$$
$$\beta_\pm^{fh} = fh_\pm B^2 + fh_\pm^{-1} D^2$$
$$\gamma_\pm^{fh} = h_\pm AB + h_\pm^{-1} CD$$

(4)

$f$ is also the refractive index of the incident medium. In our method we consider vacuum front so $f=1$.

The reflectivity, $R_\pm(q) = |r_\pm(q)|^2$, depends on $\alpha_\pm^{fh}$, $\beta_\pm^{fh}$ and $\gamma_\pm^{fh}$ in terms of a new defined quantity, $\Sigma_\pm$:

$$\sum_\pm(q) = 2\frac{1+R_\pm}{1-R_\pm} = \alpha_\pm^{fh} + \beta_\pm^{fh}$$

(5)

Alternatively, Eq. (5) can be written in the form [10]:

$$\sum_\pm = fh_\pm^{-1}\tilde{\alpha}_\pm^{ff} + h_\pm f^{-1}\tilde{\beta}_\pm^{ff} = h_\pm(A^2 + B^2) + (h_\pm)^{-1}(C^2 + D^2)$$

(6)

where the tilde denotes the mirror-reversed unknown film which is determined from interchanging the diagonal elements of the corresponding transfer matrix ($A \leftrightarrow D$), and the superscript $ff$ represents the same surrounding on both sides which in our case is vacuum.

The polarization of incident $(p_x^0, p_y^0, p_z^0)$ and reflected $(p_x, p_y, p_z)$ neutrons are given in term of $r_\pm$ and $R_\pm$ by the following relations [8]:

$$p_x + ip_y = \frac{2r_+^* r_-(p_x^0 + ip_y^0)}{R_+(1+p_z^0) + R_-(1-p_z^0)}$$

(7)

$$p_z = \frac{R_+(1+p_z^0) - R_-(1-p_z^0)}{R_+(1+p_z^0) + R_-(1-p_z^0)}$$

(8)

Using Eqs. (2) and (5)-(8), we obtain [12]:

$$\frac{p_x p_x^0 + p_y p_y^0}{p_x^{0^2} + p_y^{0^2}} = 1 + 2\frac{2-(\frac{h_+}{h_-}+\frac{h_-}{h_+}) - p_z^0(\Sigma_+ - \Sigma_-)}{\Sigma_+\Sigma_- + 2p_z^0(\Sigma_+ - \Sigma_-) - 4}$$

(9)

$$p_z = \frac{2(\Sigma_+ - \Sigma_-) + p_z^0(\Sigma_+\Sigma_- - 4)}{(\Sigma_+\Sigma_- - 4) + 2p_z^0(\Sigma_+ - \Sigma_-)}$$

(10)

Here we show how the complex reflection coefficient can be determined by using Eq. (6), Eq. (9) and two polarization measurements in one direction corresponding to two different magnetic field directions. We suppose that the experimental setup for polarization measurement of the reflected beam is arranged for the polarization in y direction.

**Step One**: suppose the incident beam to be fully polarized in the *y* direction (see figure 1). If the polarization of the reflected beam is also measured in the same direction as the incident beam, using Eqs. (6) and (9), one can easily infer that:

$$(h_+ h_-)^2 (A^2 + B^2)^2 + (C^2 + D^2)^2 + 2h_+^2(A^2 + B^2)(C^2 + D^2) = \zeta$$

(11)

where

$$\zeta = 4h_+ h_- + 2(h_+ - h_-)^2 / (1-P_1)$$

(12)

is a known parameter as a function of $P_1$ which can be determined by measuring the polarization of the reflected beam in the same direction as the polarization of the incident beam and $h = (1 - 4\pi\rho/q^2)^{0.5}$.

**Step two**: Suppose another reflectometry experiment is performed with unpolarized incident neutrons and the polarization of the reflected beam is measured in the z direction which is parallel to the spin quantization in the magnetic substrate. Measuring the polarization of the reflected beam in the z direction might bring this contradiction to mind that the polarization is being measured in a different direction than the experimental setup of the previous step. As we assume that the experimental setup to be fixed to measure the polarization of the reflected beam in y-direction, we propose to rotate the whole sample in the y-z plane by 90° and measure the polarization of the reflected beam in y direction. Considering this measured polarization as $P_2$, and using Eq. (10) we have

$$h_+ h_- (A^2 + B^2) - (C^2 + D^2) = \lambda$$

(13)

where

$$\lambda = \frac{P_2(\zeta - 4h_+h_-)}{2(h_+ - h_-)} = \frac{P_2(h_+ - h_-)}{(1 - P_1)}$$

(14)

$\lambda$ is also a known parameter from experiment as a function of $P_1$ and $P_2$.

Using Eqs. (11) and (13), the term $(A^2+B^2)$ can be determined by solving the following quadratic equation:

$$(A^2 + B^2)^2 - \frac{\lambda}{h_+h_-}(A^2 + B^2) + \frac{\lambda^2 - \zeta}{2h_+h_-(h_+h_- + h^2)} = 0$$

(15)

By knowing $(A^2+B^2)$ and using Eqs. (2), (3) and (13), the complex reflection coefficient for the free reversed unknown film can be determined as follows:

$$\tilde{r}(q) = \frac{\lambda + (1 - h_+h_-)(A^2 + B^2) \pm 2i\sqrt{\frac{\zeta - \lambda^2}{2(h_+h_- + h^2)} - 1}}{(1 - h_+h_-)(A^2 + B^2) + (2 - \lambda)}$$

(16)

Eq. (16) leads to two different possibilities. However, as it will be shown in the following example, only one of the results is physically acceptable. The physical solution is readily identified in most cases based on this fact that as $q \to 0$, $r(q) \to -1$ through negative values (predominantly positive SLD) or positive values (predominantly negative SLD) [6]. Once $r(q)$ is determined in the whole range of the $q$ values, the SLD of the sample can be retrieved using the Gel'fand-Levitan-Marchenko integral equation [2].

From experimental point of view, some considerations have to be taken in to account while one deals with the measurement of polarization in either of the two steps. In the first step, as the polarization direction of the incident neutrons and the saturating magnetic field of the substrate are not in the same direction, the external magnetic guide field would influence the saturation of the substrate. As a consequence, different magnetic domains with different orientations would be formed which would affect the reflected neutrons. In other ways, Different domains produce stray magnetic fields and as a consequence different neutrons will be reflected from the structure on optically different backing media which would lead to de-phasing. Similar concern is existed for step two. In the second step, the substrate is magnetized parallel to the external field. Applying a very strong field, one could in principle obtain a one-domain state for the large magnet. Even then the magnetic substrate would produce stray fields, which can affect the transport of polarization from the sample to the analyzer as the neutron beam passes near the pole of the large magnet. Basically, we have founded the theory of our method based on this assumption that the external guide field has no effect on the saturation of the magnetic substrate. In the next section, the method is proved to be numerically reliable. Furthermore, retrieving the SLD of the sample from the phase information of the sample is also endorsing the correctness of the method. Promisingly, we hope the method will be tested experimentally in the future and any experimental inefficiency could be resolved.

### 3. Numerical Example

In order to test the method numerically, we consider a bilayer (as unknown film) composed of 20 nm thick Au over a 30 nm thick Pt with the SLD of 4.66 and 6.34 $\times 10^{-4}$ nm$^{-2}$ for Au and Pt respectively. As shown in Fig. 2, it is supposed that the sample is mounted on top of a Co substrate with nuclear SLD of 2.26$\times 10^{-4}$ nm$^{-2}$ and magnetic SLD of $\pm 4.12\times 10^{-4}$ nm$^{-2}$ for incident neutrons parallel (+) and antiparallel (−) to the magnetization of the of the magnetic substrate. In order to determine the term $(A^2+B^2)$, two different measurements of the reflected neutrons ($P_1$, $P_2$) have to be performed in the same order as was explained as step one and two in the previous section. Knowing $\zeta$ and $\lambda$ from measurement of $P_1$ and $P_2$ and solving the quadratic equation of (16), the term $(A^2+B^2)$ is obtained.

Fig.4, illustrates the real part of reflection coefficient, Re$r(q)$, for mirror reversed image of the sample of Fig.2 with vacuum surroundings on both sides . The curve with blue circles represents the recovered data of Re$r(q)$ for larger values than critical $q$ of the substrate. The data bellow the critical $q$ can readily be interpolated from the fact that Re$(r) \to -1$ as $q \to 0$ [6]. For better illustration of the results, q$^2$Re(r) is plotted in Fig. 4.

The imaginary part of the reflection coefficient for the mirror image of the sample with vacuum surroundings on both sides, Im$r(q)$, is also determined by knowing $(A^2+B^2)$. Since Eq. (16) would lead to two different solutions for the imaginary part of the reflection coefficient, the physical solution is chosen based on this fact that as $q \to 0$, $Im(r) \to 0$ through negative values (predominantly positive SLD) or positive values (predominantly negative SLD). As it is shown in Fig.5, the circles and squares show the two different solution for Im$r(q)$. The physical branch alternates between these sets.

In order to retrieve the scattering length density of the sample, we use the real and imaginary parts of the reflection coefficient as input in some certain codes such as the one which is developed by P. Sacks [14,15]. Fig. 6, demonstrates the retrieved SLD for the mirror image of the sample of Fig. 2 with vacuum surroundings on both sides. As it is illustrated in the figure, the retrieved SLD for the mirror image of the sample of Fig.1 with vacuum surroundings on both sides is clear. The whole results certify that the method is applicable for available reflectometers and the complex reflection coefficient of the reversed unknown sample can be determined by one directional polarization analysis of the reflected beam.

## 4. Conclusion

We have proposed a method to determine the modules and phase of reflection coefficient by using a magnetic substrate and polarization measurement. The method takes into account the limitation of available reflectometers in polarization analysis in which the polarization measurements of reflected beam can only be performed in the same direction as the incident beam. In comparison with other methods which use polarization measurements, this new approach requires just two measurement of polarization in one direction (y-direction as normal to the plane of reflection) corresponding to one magnetic field in two different directions. One of the polarization measurements is performed with incident beam polarized in y-direction and the other for unpolarized incident beam while the whole sample and its quantizing magnetic field are rotated in the y-z plane by 90°.

In this new method, no reflectivity data are required and the relations are more simple and straightforward. The same as the standard method of variation of surroundings [5,6], the scheme provides us with information of phase and modules of the reflection coefficient for the mirror image of the sample with vacuum surroundings at both side. Finally, the retrieved information of the reflection coefficient in the whole range of $q$ values makes it possible to reconstruct a unique profile for the sample.

In summary a new method to solve the phase problem of neutron reflectometry has been worked out in principle and tested numerically. The problems envisaged in realization of the method are discussed and its experimental implementation is a challenge to be taken up.

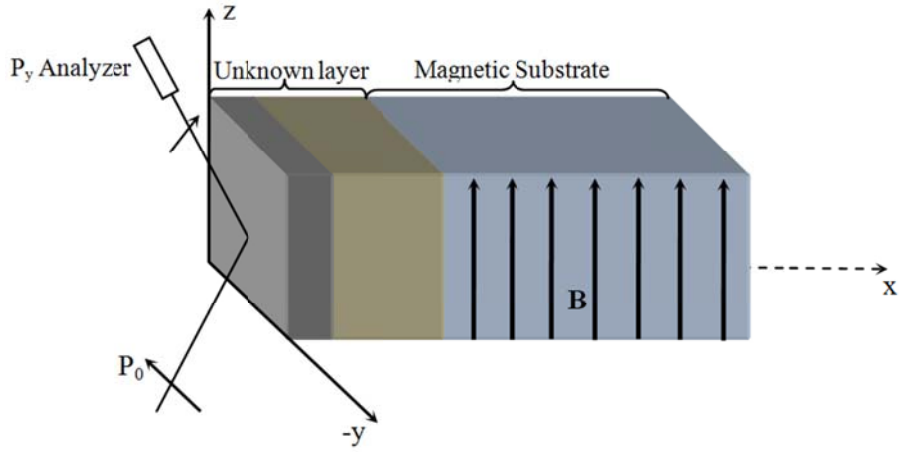

Figure 1: The schematic illustration of the coordinate system, layers and the polarization of the beam.

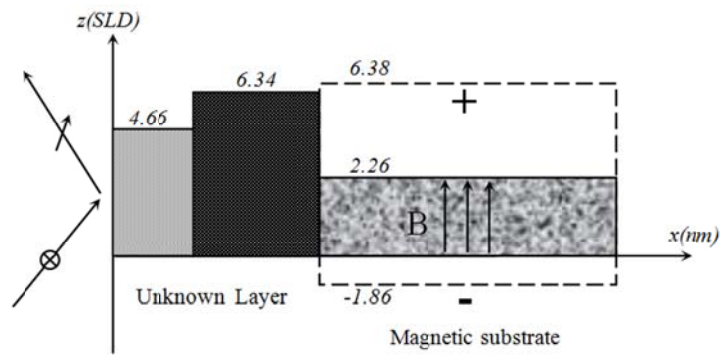

**Figure 2:** The SLD of the sample used for numerical example. Dashed line represents the effective potential experienced by neutrons due to the presence of magnetic field, **B,** inside the substrate. The crossed circle denotes the polarization direction of the incident beam in step one measurement.

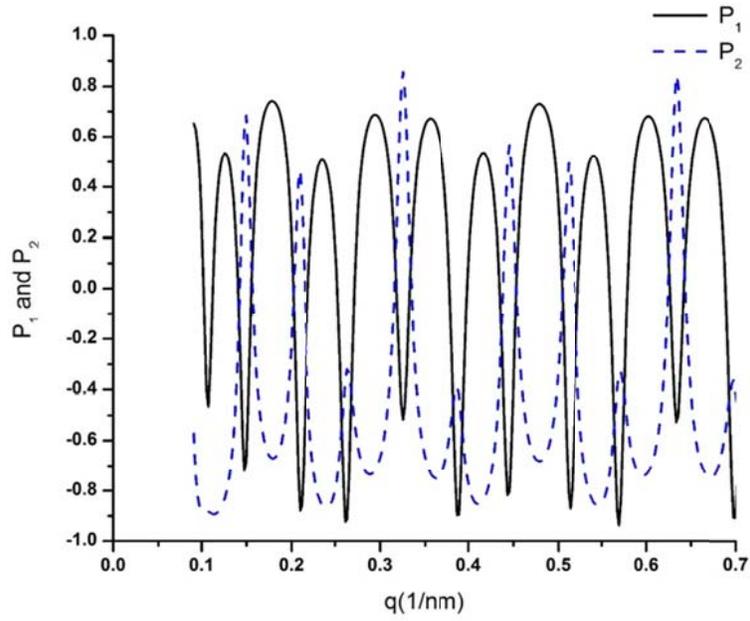

**Figure 3**: The polarization of the reflected beam in y-direction for two reflectometry experiments outlined in section 2.

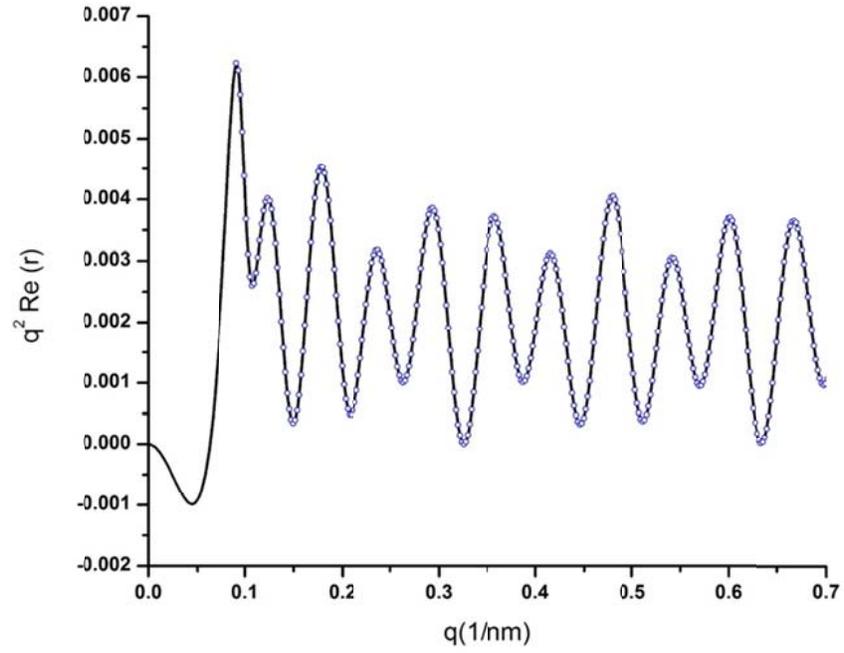

**Figure 4:** The real part of the reflection coefficient for the mirror image of the film in Fig. 1 without backing. Solid line: Calculated directly by Eq. (2). Circles: Recovered for larger values than critical $q$ of the substrate, from the two measurement of $P_y$ corresponding to magnetic fields in two different direction

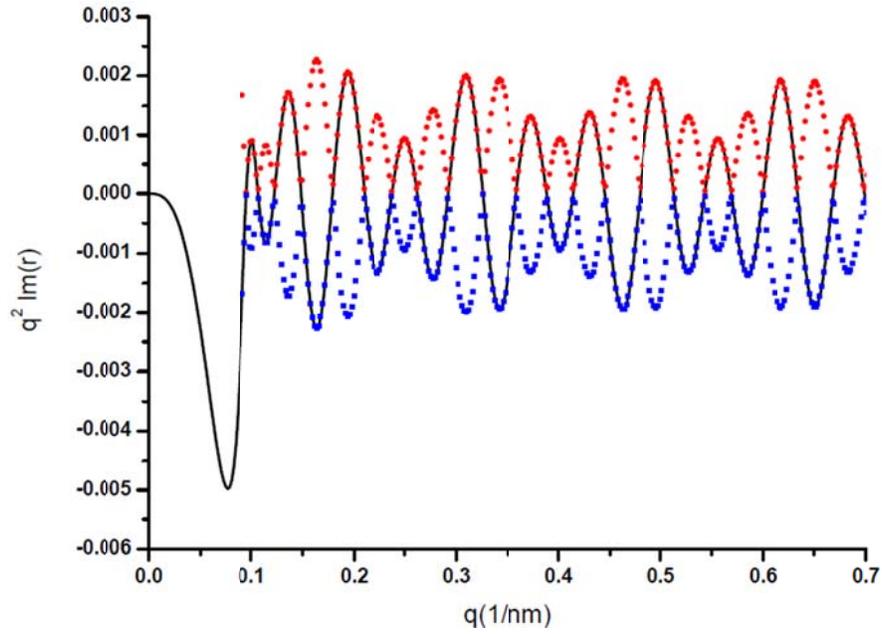

**Figure 5:** The imaginary part of the reflection coefficient for the mirror image of the film in Fig. 1 without backing. Solid line: Calculated directly by Eq. (2). Circles and squares: Recovered for larger values than critical $q$ of the substrate, from the two measurement of $P_y$ corresponding to magnetic fields in two different directions. The physical branch alternates between these sets.

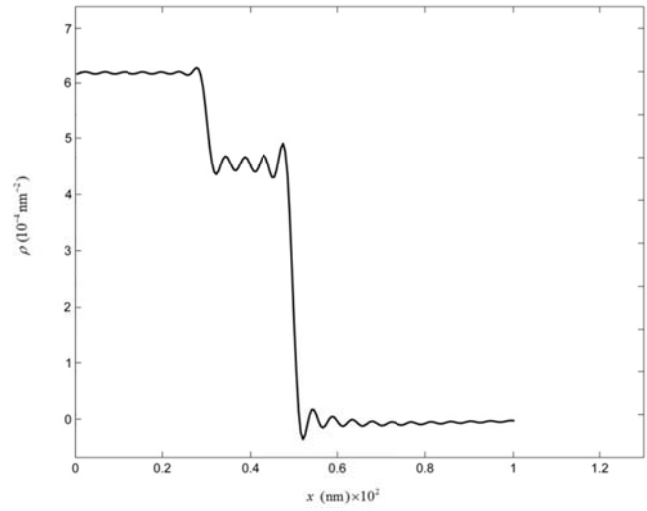

**Figure 6:** The retrieved SLD for the mirror image of the sample of Fig.2 without backing using the recovered data of Figs. 4 and 5 and the Sacks Code [14, 15].